\documentclass[a4paper,10pt]{article}
\usepackage{amsmath}
\usepackage{graphicx,psfrag,epsf}
\usepackage{enumerate}
\usepackage{natbib}

\newcommand{\blind}{0}

\usepackage{algorithmic}
\usepackage{algorithm}
\usepackage{epsfig,amssymb,amsmath,amsthm,xspace,rotating}
\bibpunct{[}{]}{,}{a}{}{;}
\numberwithin{equation}{section}
\numberwithin{figure}{section}
\theoremstyle{plain}
\newtheorem{theorem}{Theorem} %[section]
\newtheorem{corollary}[theorem]{Corollary}

\newtheorem{lemma}{Lemma}
      
\theoremstyle{definition}
\newtheorem{definition}{Definition}[section]
\newtheorem*{notation*}{Notation}
\newtheorem*{example*}{Example}
      
\theoremstyle{remark}
\newtheorem{remark}{Remark}[section]

\newcommand{\be}{\begin{equation*}}
\newcommand{\ee}{\end{equation*}}
\newcommand{\bea}{\begin{align*}}
\newcommand{\eea}{\end{align*}}
\newcommand{\R}{{\mathbb R}}

\newcommand{\norm}[1]{\ensuremath{\left\Vert #1 \right\Vert}\xspace}

\newcommand{\haf}[1]{\ensuremath{\widehat{f}_{ #1 }}\xspace}

\newcommand{\seq}[2]{\ensuremath{#1, \ldots, #2}}
\newcommand{\sign}{\ensuremath{\mathrm{sign}}}
\allowdisplaybreaks
\begin{document}

\def\spacingset#1{\renewcommand{\baselinestretch}%
{#1}\small\normalsize} \spacingset{1}

%%%%%%%%%%%%%%%%%%%%%%%%%%%%%%%%%%%%%%%%%%%%%%%%%%%%%%%%%%%%%%%%%%%%%%%%%%%%%%

\if0\blind
{
  \title{\bf LASSO ISOtone for High Dimensional Additive Isotonic Regression}
  \author{Zhou Fang\hspace{.2cm}\\
    Department of Statistics, University of Oxford\\
    and \\
    Nicolai Meinshausen \\
    Department of Statistics, University of Oxford}
  \maketitle
} \fi

\if1\blind
{
  \bigskip
  \bigskip
  \bigskip
  \begin{center}
    {\LARGE\bf LASSO ISOtone for High Dimensional Additive Isotonic Regression}
\end{center}
  \medskip
} \fi

\bigskip
\begin{abstract}
Additive isotonic regression attempts to determine the relationship between a multi-dimensional observation variable and a response, under the constraint that the estimate is the additive sum of univariate component effects that are monotonically increasing. In this article, we present a new method for such regression called LASSO Isotone (LISO). LISO adapts ideas from sparse linear modelling to additive isotonic regression. Thus, it is viable in many situations with high dimensional predictor variables, where selection of significant versus insignificant variables are required. We suggest an algorithm involving a modification of the backfitting algorithm CPAV. We give a numerical convergence result, and finally examine some of its properties through simulations. We also suggest some possible extensions that improve performance, and allow calculation to be carried out when the direction of the monotonicity is unknown.
\end{abstract}

\noindent%

\spacingset{1.45}

\section{Introduction}
\label{chap:intro}

We often seek to uncover or describe the dependence of a response on a large number of covariates. In many cases, parametric and in particular linear models may prove overly restrictive. Additive modelling, as described, for instance in \citet{hastieadd}, is well known to be an useful generalisation.

Suppose we have $n$ observations available of the pair $\left(X_i, Y_i\right)$, where $Y_i \in \R$ is a response variable, and $X_i = ( X_i^{(1)}, \ldots, X_i^{(p)}) \in \R^{p}$ is a vector of covariates.

In additive modelling, we typically assume that the data is well approximated by a model of the form
\[Y_i = \sum_{k=1}^p f_k(X^{(k)}_i) + \varepsilon_i,\]

where $\varepsilon = (\seq{\varepsilon_1}{\varepsilon_n})$ is a random error term, assumed independent of the covariates and identically distributed with mean zero. For every covariate $k = \seq{1}{p}$, each component fit $f_k$ is chosen from a space of univariate functions $\mathcal{F}_k$. Usually, these spaces are constrained to be smooth in some suitable sense, and in fitting, we minimise the L2 norm of the error,

\[\frac{1}{2}\norm{Y - \sum_{k=1}^p f_k(X^{(k)})}^2 := \frac{1}{2}\sum_{i=1}^n \left(Y_i - \sum_{k=1}^n f_k(X^{(k)}_i) \right)^2,\]

under the constraint that $f_k \in \mathcal{F}_k$, for each $k = \seq{1}{p}$. In the case that $\varepsilon$ is assumed to be normal, this can be directly justified as maximising the likelihood.

Work on such methods of additive modelling have produced a profuse array of techniques and generalisations. In particular, \citet{addiso} suggested the additive isotonic model. With the additive isotonic model, we are interested in tackling the problem of conducting regression under the restriction that the regression function is of a pre-specified monotonicity with respect to each covariate. (Isotonic means the functions are increasing, though decreasing can be accommodated easily by reversing the signs of covariates.) Such restrictions may be sensible whenever there is subject knowledge about the possible influence or relationship between predictor and response variables. A broad survey of the subject may be found in \citet{underorder}. It turns out that in the univariate case, the Pool Adjacent Violators Algorithm, as first suggested in \citet{pava:1955}, allows rapid calculation of a solution to the least squares problem using this restriction alone. By doing so, we retain only the ordinal information in the covariates, and hence obtain a result that is invariant under strictly monotone transformations of the data. In addition, the form of the regression, being simply a maximisation of the likelihood, means that apart from the monotonicity constraint, we do not put on any regularisation, or smoothing.

\citet{addiso} built on this, by generalizing to multiple covariates. Here, the regression function is considered to be a sum of univariate functions of specified monotonicity. Fitting is conducted via the cyclic pool adjacent violators (CPAV) algorithm, in the style of a backfitting procedure built around PAVA --- that is, cycling over the covariates, the partial residuals using the remaining covariates are repeatedly fitted to the current one, until convergence. Later theoretical discussion from \citet{addisomammen} outlined some positive properties of this procedure.

Nevertheless, CPAV, like many types of additive modelling, can fail in the high dimensional case --- for instance, once $p > n$. The particular problem is that the least squares criterion loses strictness of convexity when the number of covariates is large, since it becomes easy for allowed component fits in some covariates to combine in the training data so as to replicate component fits in unrelated covariates. It is hence impossible for the CPAV to distinguish between two radical different regression functions since they give the same fitted values on the training dataset. Some success might be achieved, though, if the solution sought is sparse, in the sense that most of the covariates have little or no effect on the response. If the identity of the significant variables were known, then, the CPAV could be conducted on a much smaller set of covariates. However, exhaustive search to identify this sparsity pattern would be rapidly prohibitive in terms of computational cost, scaling exponentially in the number of covariates.

In the context of parametric linear regression, it has emerged recently that such sparse regression problems can be dealt with by use of a L1-norm based penalty in the optimisation. This can resolve the identifiability problem and achieve good predictive accuracy. \citet{tibLASSO, donoho}, amongst others, have identified several significant empirical and theoretical results to support this `LASSO' estimator, while \citet{efronlars, pathwisecoord} and others have invented fast algorithms for calculating both individual estimates and full LASSO solution paths.

Generalisation of the L1 penalisation principle to nonparametric regression can also lead to successful with additive modelling. For example, recent work on this subject includes SpAM \citep{ravispam}, which describes the application of the grouped LASSO to general smoothers, and high dimensional additive modelling with smoothness penalties \citep{lukasham} which follows similar principles, using a spline basis.

In this paper, we propose the Lasso-Isotone (LISO) estimator. By modifying the additive isotonic model to include a LASSO-style penalty on the total variation of component fits, we hope to conduct isotonic regression in the sparse additive setting. 

The LISO is similar to the degree $0$ case of the LASSO knot selection of \citet{osbknot}, which is also identical to the fused LASSO of \citet{tibfused}, if we replace the covariate matrix with ordered Haar wavelet bases, and do not consider coefficient differences for coefficients corresponding to different covariates. It is also similar to the univariate problem considered by \citet{localadapt}. In contrast to each of these procedures, however, we allow the additional imposition of a monotonicity constraint, producing an algorithm similar in complexity to the CPAV.

In section~\ref{methodsec} we shall describe the LISO optimisation, and in section~\ref{algsec} we will discuss algorithms for computation for fairly large $n$ and $p$. We will discuss the effect of the regularisation, and then in section~\ref{sec:ext} suggest some extensions. Finally, in section~\ref{simusec} we will explore its performance using some simulation studies. Proofs of theorems are left for the appendix.

\section{The LASSO-ISOtone Optimisation}\label{methodsec}

For now, let us assume without loss of generality that we are conducting regression constrained to monotonically increasing regression functions. Let us first define some terms.

Let $Y \in \R^n$ be the response vector. Assume, subtracting by a constant intercept term if necessary, that $\sum_{i=1}^n Y_i = 0$. $X = \left(\seq{X^{(1)}}{X^{(p)}}\right) \in \R^{n \times p}$ is the matrix of covariates. 

For a specified $X$, for $k = \seq{1}{p}$, let  $\mathcal{F}_{k}$ be the space of bounded, univariate, and monotonically increasing functions, that have expectation zero on the $k$-th covariate. $-\mathcal{F}_{k}$ is then the same for monotonically decreasing functions.

\begin{align*}
\mathcal{F}_{k} :=\left\{f: \R \rightarrow \R \,\left\vert \, \sum_{i=1}^n f\left(X^{(k)}_i\right) = 0, \textrm{ and } \exists\, U, V \textrm{ s.t. } \forall a < b, U \leq f(a) \leq f(b) \leq V  \right.\right\} 
\end{align*}

Additive isotonic models involve sums of functions from these spaces. It is simple to observe that each $\mathcal{F}_{k}$ is a convex half-space that is closed except at infinity, and so as a result the space of sums of these functions must also be convex and closed except at infinity.

\begin{definition}
We define the Lasso-Isotone (LISO) solution for a particular value of tuning parameter $\lambda \geq 0$ as the minimiser $\widehat{f}_\lambda = \left( \widehat{f}_{k, \lambda} \right)_{k=1}^p$, with $\widehat{f}_{k, \lambda} \in \mathcal{F}_{k} \, \forall k$, of the LISO loss
\begin{equation}
L_\lambda \left( \seq{f_1}{f_p} \right) := \frac{1}{2} \norm{Y - \sum_{k=1}^p f_k\left(X^{(k)}\right)}^2 + \lambda \sum_{k=1}^{p} \Delta(f_k). \label{lambdaop}
\end{equation}
\end{definition}

Here $\Delta(f_k)$ denotes the total variation of $f_k$, which for $f_k \in \mathcal{F}_{k}$ can be calculated as

\[\Delta(g) = \sup_{x\in \R} f_k(x) - \inf_{x \in \R} f_k(x).\]

As with the LASSO, the LISO objective function is the sum of a log-likelihood term and a penalty term. It is clear that the domain is convex and, considered in the space of allowed solutions, the objective itself is convex and bounded below. Indeed, outside a neighbourhood of the origin, both terms in the objective are increasing, so a bounded solution exists for all values of $\lambda$. However, the objective may not be strictly convex, so this solution may not be unique.

The log-likelihood term does not consider the values of $f_k$ except at observed values of each covariate, while the total variation penalty term, assuming monotonicity, only takes account of the upper and lower bounds of the covariate-wise regression function --- indeed, for optimality, these bounds must be attained at the extremal observed values of the appropriate covariate, with the solution flat beyond this region. Thus, given any one minimiser to $L_\lambda$, another fit with the same function values at observed covariate points, interpolating monotonically between them, will have the same value of $L_\lambda$, and so also be a LISO solution. This means that the we can equivalently consider optimisation in the finite dimensional space of fitted values $\widehat{f}_k(X^{(k)})$.

For simplicity, we will represent found LISO solution components by the corresponding right continuous step function with knots only at each observation. For the remainder of this paper, we shall consider uniqueness and equivalence in terms of having equal values at the observed $X^{(k)}$.

We have introduced a mean zero constraint on the fitted components for identifiability, since we can easily add a constant term to any component fit $f_k$, and deduct it from another component, and still arrive at the same final regression function. However, we will show later that this constraint arises naturally in the univariate case, where even without it being explicitly applied,
\[\sum_{i=1}^{n}\widehat{f}(X_i) = \sum_{i=1}^{n}Y_i.\]

The total variation penalty shown here has been previously suggested for regression in \citet{localadapt}, though in that case, the focus was on smoothing of univariate functions, without a monotonicity constraint.
% short version
% \subsection{Generic LASSO algorithms and LISO-LARS}\label{genericLASSO}
\section{LISO Backfitting}\label{algsec}

Considering the representation of the LISO in terms of step functions, the LISO optimisation for a given dataset can be viewed as ordinary LASSO optimisation for a linear model, constrained to positive coefficients, using an expanded design matrix $\widetilde{X} \in \R^{n \times p(n-1)}$, where $\widetilde{X} = \left(\widetilde{X}^{(1)}\, \ldots \, \widetilde{X}^{(p)}\right)$. Each $\widetilde{X}^{(k)} \in \R^{n \times (n-1)}, \quad k = 1, \ldots, p$ contains $n-1$ step functions in the $k$-th covariate, which form a basis for the vector $f_k(X^{(k)})$, and so isotonic functions in that covariate. The coefficients $\beta$ optimised over then represent step sizes.

Such a construction is suggested in \citet{osbknot}, amongst others. Under this re-parametrisation of the problem, existing LASSO algorithms for linear regression may be applied, with a modification to restrict solutions to non-negative values. In particular, the Least Angle Regression algorithm of \citet{efronlars} is effective, since shortcuts exist for calculating the necessary correlations.

On the other hand, the high dimensionality of $\widetilde{X}$ means that standard methods become very costly in higher dimensions, both in terms of required computation, but especially in terms of the storage requirements associated with very large matrices. Hence, we must consider more specialised algorithms for such cases. One such approach involves backfitting, and is workable due to the simple form of the solution when restricted to a single covariate.

\subsection{Thresholded PAVA}

In the $p=1$ case, it turns out that we have an exceptionally simple way to calculate the LISO estimate, which we will later use to establish a more general multivariate procedure.

With no LISO penalty (i.e. $\lambda = 0$) and a single covariate, the LISO optimisation is equivalent to the standard univariate isotonic regression problem. In this case, the loglikelihood residual sum of squares term is strictly convex, and so, as a strictly convex optimisation on a convex set, an unique solution exists. Trivially, the solution must also be bounded. In fact, there exists, as described in \citet{underorder} and attributed to \citet{pava:1955}, a fast algorithm for calculating the solution -- the Pool Adjacent Violators Algorithm (PAVA).

Hence, defining $\haf{\lambda}$ as the solution to optimisation~\eqref{lambdaop} for $\lambda$, we have $\haf{0} = \widehat{f}_{\mathit{PAVA}}$ . The following theorems describe the solutions for other values of $\lambda$:

\begin{theorem}\label{thresh1} 
For $A \leq B$, denote by \haf{>A, <B} the Winsorized PAVA estimate
\begin{equation*}
\haf{>A, <B}(x) := \begin{cases}
A & \textrm{if }\haf{\mathit{PAVA}}(x) < A\\
B & \textrm{if }\haf{\mathit{PAVA}}(x) > B\\
\widehat{f}_{\mathit{PAVA}}(x) & \mathrm{otherwise}.
\end{cases}
\end{equation*}

Then if $p=1$, there exist thresholds $A_\lambda \leq B_\lambda$ for each value of $\lambda \geq 0$ such that the LASSO-Isotone solution is given by $\widehat{f}_\lambda \equiv \haf{>A_\lambda, <B_\lambda}$.
\end{theorem}

\begin{theorem}\label{threshfind}
In Theorem~\ref{thresh1}, given $\haf{\mathit{PAVA}}$, the pair $A_\lambda, B_\lambda$ (the optimal thresholding levels) are a piecewise linear, continuous and monotone (increasing for $A_\lambda$, decreasing for $B_\lambda$) function of $\lambda$, for $\lambda\geq 0 $.

Specifically, if 
\begin{equation}
2\lambda \geq \sum_{i=1}^n \vert \haf{\mathit{PAVA}}(X_i) - \overline{Y} \vert, \label{ifzero2}
\end{equation}
then $A_\lambda = B_\lambda = \overline{Y}$.

Otherwise, $A_\lambda, B_\lambda$ are the solutions to
\begin{align}
\sum_{i =1}^n (\haf{\mathit{PAVA}}(X_i) - B_\lambda)_+ &= \lambda \label{Bmini}\\
\sum_{i =1}^n (A_\lambda - \haf{\mathit{PAVA}}(X_i))_+ &= \lambda.\label{Amini}
\end{align}
\end{theorem}
\begin{corollary}\label{otherzero}
Let $\pi$ be a permutation taking $1, \ldots, n$ to indices that put $X$ in ascending order. Then if
\begin{equation}
\lambda \geq \max_{m} \left\vert \sum_{i=1}^m \left( Y_{\pi(i)} - \overline{Y} \right)  \right\vert 
\end{equation}
$A_\lambda = B_\lambda = \overline{Y}$.
\end{corollary}

\begin{remark}
The LHS of \eqref{Bmini} and \eqref{Amini} specify the amount by which each threshold changes the sum of the fitted values on the appropriate side of the mean. Hence, we see that $\sum_{i=1}^n\haf{\lambda}(X_i) = \sum_{i=1}^n \haf{\mathit{PAVA}}(X_i) = \sum_{i=1}^n Y_i$, for all $\lambda$.

In other words, if $Y$ has mean zero, then the mean zero constraint on the fit arises naturally, without having to be externally applied. If $Y$ does not have mean zero, the solution is simply a shifted version of the fit for $Y - \overline{Y}$. This justifies deducting the mean of the response and dealing with it separately.

\end{remark}
\begin{remark}
The PAVA algorithm itself can accommodate observation weights, as well as tied values in the covariates. In terms of the LISO, working with unequal observation weights demands that we work with weighted residual sums of squares. This does not affect Theorem~\ref{thresh1}, but for equations~\eqref{Amini} and \eqref{Bmini}, weights should be introduced in the summation. Tied values should be also dealt with by merging the relevant steps, and weighting them according to the number of data points at that covariate observation.
\end{remark}

\subsection{Backfitting algorithm}
In general, however, simple thresholding fails to solve the LISO optimisation in higher dimensions, due to correlations between steps in different covariate component functions. We can, however, extend the 1D algorithm to higher dimensions by applying it iteratively as a backfitting algorithm.

In other words, we define LISO-backfitting by the following steps:

\begin{algorithm}
  \caption{LISO-Backfitting}
  \label{alg:LISOback}
  \begin{algorithmic}[1]
    \STATE Set $m = 0$.
    \STATE Initialise component fits $\left(\seq{f_1}{f_p}\right)$ as identically 0, or as the estimate for a different value of $\lambda$, storing these as the $n \times p$ marginal fitted values. 
\REPEAT
    \STATE $f^m \Leftarrow \left(\seq{f_1}{f_p}\right)$.
    \STATE $m \Leftarrow m + 1$.
    \FOR {$k = 1$ to $p$ (or a random permutation)}
	\STATE Recalculate residuals $r_i \Leftarrow Y_i - \sum_{k=1}^p f_k\left(X^{(k)}_i\right), \, i = 1,\ldots, n$.
    \STATE Refit conditional residual $\left\lbrace r_i + f_k\left(X^{(k)}_i\right) \right\rbrace_{i=1}^n$ using $X^{(k)}$ by PAVA, producing $\widetilde{f}_k\left(X^{(k)}_i\right)$, for $i=1,\ldots,n$.
\STATE Calculate thresholds $A_\lambda$, $B_\lambda$ from $\lambda$ and $\widetilde{f}_k$ by Theorem~\ref{threshfind}.
\STATE Adjust component fit $f_k(X^{(k)}_i) \Leftarrow \widetilde{f}_{k, >A_\lambda <B_\lambda}(X^{(k)}_i)$.
    \ENDFOR
\UNTIL{sufficient convergence is achieved, through considering $f^m$ and $f^{m-1}$.}
\STATE Interpolate $f_k$ between the samples $X_i^{(k)}$.
  \end{algorithmic}
\end{algorithm}

\begin{theorem}\label{converges} For $f^m = \left(f^m_1, \ldots, f^m_p\right)$, the sequence of states resulting from the LISO-backfitting algorithm, $L_\lambda(f^m)$ converges to its global minimum with probability 1. Specifically, if there exists an unique solution to \eqref{lambdaop}, $f^m$ converges to it.
\end{theorem}

\begin{remark}\label{remarkconv} 
If there is no unique solution, the backfitting algorithm may not necessarily converge, though the LISO loss of each estimate will converge monotonically to the minimum. In addition, because the objective function is locally quadratic, as the change in the LISO loss converges to zero, the change in the estimate after each individual refitting cycle converges also to zero.
\end{remark}
\begin{remark}
Moreover, defining $X^{(k)}_{(i)}$ as the $i$-th smallest value of $X^{(k)}$, if a certain individual step in the final functional fit \[f_k\left(X^{(k)}_{(i)}\right) - f_k\left(X^{(k)}_{(i-1)}\right)\] has a value of zero in all solutions to the LISO minimisation, then, after a finite number of steps, all results from the algorithm must take that step exactly to zero.

This is because steps being estimated as zero in a LISO solution implies that the partial derivative of the LISO objective function $L_\lambda$ in the above individual step direction is greater than zero when evaluated at this solution. The partial derivatives are continuous, so as the algorithm converges, the partial derivatives associated with zero steps eventually be above 0 and remain so. But then, this can only be the case following a thresholded PAVA calculation involving the covariate associated with that step if that single covariate optimisation takes the step exactly to zero.
\end{remark}

Convergence of the algorithm can be checked for by a variety of methods. One of the simplest is to note that due to the nature of the repeated optimisation, the LISO loss will always decrease in each step, and we will converge towards the minimum. Hence, one viable stopping rule would be to cease calculating when the LISO loss of the current solution drops by too small an amount. Alternatively, we can exploit Remark~\ref{remarkconv}, and monitor the change in the results in each cycle, stopping when this becomes small.

\subsection{Choice of regularisation parameter}

It will be always necessary to choose a tuning parameter $\lambda$ to facilitate appropriate fitting. As with the LASSO, too high a tuning parameter will shrink the fits towards zero. Indeed, consideration of Corollary~\ref{otherzero} shows that, with $\overline{Y} = 0$, and $\pi^{(k)}$ defined as a permutation that puts the $k$-th covariate into ascending order, a choice of $\lambda$ greater than
\[ \max_{\substack{k = 1, \ldots, p,\\ m = \seq{1}{n}}} \left\vert \sum_{i=1}^m Y_{\pi^{(k)}(i)} \right\vert \]
will result in a zero fit in every thresholded PAVA step starting from zero, and hence a zero fit overall for the LISO. 

Conversely, too small a value of $\lambda$ will lead to improper fitting. This arises from two sources. Firstly, as with the LASSO, the noise term may flood the fit, as the level of thresholding is not sufficient to suppress correlations of the noise with the covariate step functions -- the columns of $\widetilde{X}$. Secondly, $\lambda$ has a role in terms of fit complexity, with a small value of $\lambda$ implying that the LISO, when restricted to the true covariates, would select more steps. This means a less sparse signal in the implied LASSO problem, so it becomes in turn more likely for selected columns of $\widetilde{X}$ to be correlated with columns belonging to irrelevant covariates, hence producing spurious fits in the other covariates.

More precisely, in the noiseless case, if the true model function can be written exactly as the sum of step functions with, in the expanded design matrix $\widetilde{X}$, corresponding column indices $S$, then correct recovery, given that LISO has fit non-zero fits to the true step functions, requires

\begin{align}
\lambda &\geq \widetilde{X}_{S^c}^T\left(Y - \widetilde{X}_S\left(\widetilde{X}_S^T\widetilde{X}_S\right)^{-1}(\widetilde{X}_S^TY - \lambda)\right) \\
&=\widetilde{X}_{S^c}^T\left(\widetilde{X}_S\left(\widetilde{X}_S^T\widetilde{X}_S\right)^{-1}\lambda\right).
\end{align}

This is the Irrepresentable Condition of the LASSO, as detailed in \citet{consistLASSO,meingraph}, and it may fail if $S$ is too large. With the LISO, then, the particular choice of $\lambda$ itself influences the form the true covariates can take and so alters the criterion for Irrepresentability.

%TODO more from here
\begin{figure}[ht]
\vspace{-10pt}
 \centering
 \includegraphics[scale = 0.5]{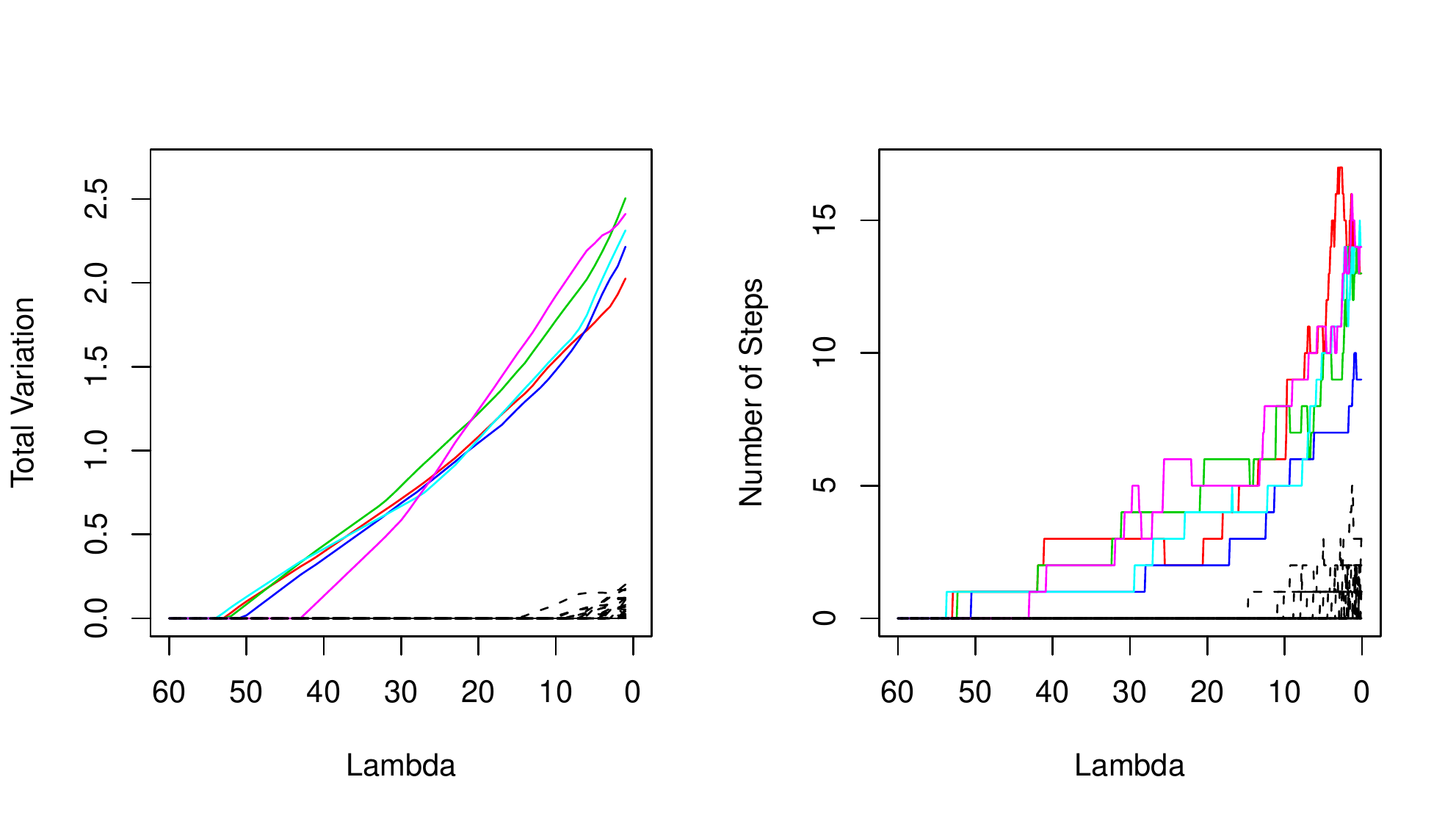}
\vspace{-10pt}
 \caption{Effects of changing the regularisation parameter in the noiseless case. $n=100, p=200$. Each line represents how an individual covariate's estimate changes as $\lambda$ varies, with the solid lines for the true covariates, while the dashed lines denote spurious fits on irrelevant variables.}
 \label{fig:fitsexample}
\end{figure}

These effects are illustrated in Figure~\ref{fig:fitsexample}, in which we have generated $X$, with $n = 100$, p=$200$, according to an uniform distribution, and produced $Y$ as the sum of $k=5$ of the covariates. In other words, $f$ is the sparse sum of linear functions. We give the full paths of fits in terms of, firstly, the total variation of fitted components $\Delta(f_k)$, and secondly the number of component steps in each covariate, 
 \[ \left\vert \left\lbrace i :  f_k\left(X^{(k)}_{(i)}\right) \neq f_k\left(X^{(k)}_{(i-1)}\right) \right\rbrace \right\vert.  \]

Of particular note is that, unlike the LASSO, even without noise, the size of the basis of step functions and the non-sparsity of the true signal means that as $\lambda \rightarrow 0$, we do not converge to the true sparsity pattern. However, with higher $\lambda$, the number of steps we choose diminishes rapidly, and as a result we can remove the spurious fits and simultaneously not mistakenly estimate the relevant covariates as zero.

\begin{figure}[ht]
\vspace{-10pt}
 \centering
 \includegraphics[scale = 0.5]{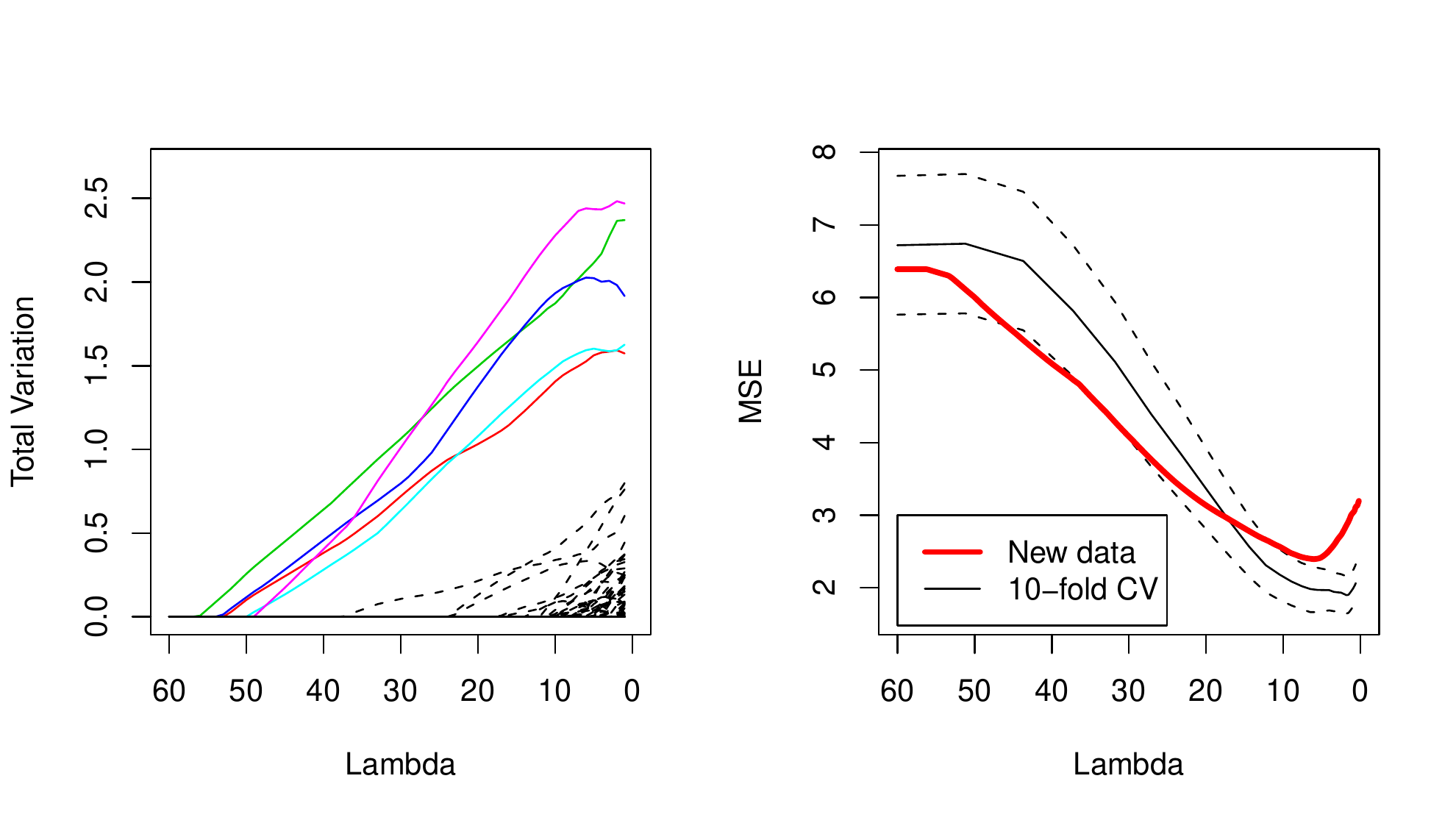}
\vspace{-10pt}
 \caption{Effects of changing the regularisation parameter in the noisy case. $n=100, p=200, SNR=5$. We show again in the first graph the total variation of each covariate estimate as $\lambda$ alters, with solid lines for the truly important covariates, while the dashed lines denote spurious fits on irrelevant variables. The second graph shows the MSE from a 10-fold cross validation procedure with $\pm 1$ s.d. in dashes, as well as the true MSE on a new set of data as the thick line.}
 \label{fig:fitsexample2}
\end{figure}

In Figure~\ref{fig:fitsexample2}, we add an independent normal noise component to $Y$, with variance chosen so that the signal to noise ratio, $SNR = 5$. In the new Total Variation plot, we see that the noise component has added additional noise fits in some of the irrelevant variables, and as in the LASSO these vanish for higher $\lambda$. Since the spurious fits vanish before the true covariate components do, we see that recovery of the true sparsity pattern is still possible in this case.

Now, in the above examples, we worked with the true sparsity pattern being assumed known. In real problems, we need to estimate the correct value of $\lambda$ directly from the data. To do this, with the goal of recovering the correct sparsity pattern, is generally understood to be very difficult. (See e.g. \citet{meingraph} for some attempts.) However, as suggested in literature from \citet{tibLASSO} onwards, cross validation is effective for minimising predictive error, and is illustrated by second graph of Figure~\ref{fig:fitsexample2}. Here, we calculate CV error from a 10-fold cross validation. We may then take the $\lambda$ that minimises the average mean squared error across the folds. If we desire a simpler model, we can, as is often suggested, take the largest $\lambda$ that achieves a CV value within 1 s.d. of the minimum. Examining the thick line for the true predictive MSE shows that such a procedure, while not perfect, can give good results. In minimising predictive error, however, we do still fit some irrelevant covariates as non-zero, a phenomenon previously observed with the LASSO in \citet{lengirrelevants}.

Now, unlike a LARS-like approach, LISO Backfitting will only give us the solution for an individual choice of $\lambda$. However, CV can still be practical, because coordinatewise minimisation can be very fast for sparse problems, something already observed for the normal LASSO \citep{pathwisecoord}. We can further reduce the computational cost by noting that LISO solutions for similar values of $\lambda$ are likely to be similar, and hence use the result for one value of $\lambda$ as a start point for the calculation for a nearby value of tuning parameter. This is especially effective if we order the $\lambda$ values we need to calculate in decreasing order, since large $\lambda$ solutions are more sparse and so faster to calculate.

\section{Extensions and variations}\label{sec:ext}

A variety of extensions and variations of the basic LISO procedure may be proposed, that may offer improvements in some circumstances.

\subsection{Bagged LISO}
Bagging \citep{bagging} may be used with the LISO, by aggregating the results of applying the LISO to a number of bootstrap samples through any of a variety of methods. This usually succeeds in smoothing the observation, especially if we use smoothed bagging \citep{smoothbag}. However, this method is not reliably a great improvement in our empirical studies. Further, since the aggregated fit will produce a sparsity pattern involving a set of selected covariates that is the union of the selected covariates for each individual subsample calculation, we have that bagging will almost inevitably reduce the degree of sparsity in the fit, for any given degree of regularisation.

\subsection{Adaptive LISO}
A potential problem with the LISO is that it treats the constituent steps of each fit individually. In other words, there is no difference, in the eyes of the optimisation, between a fit that involves single step fits in a large number of covariates, and a single more complex fit in one covariate. As a result, the method may not achieve a great deal of sparsity in terms of covariates used, an issue we may want to rectify through making the algorithm in some sense recognise the natural grouping of steps in the step function basis.

Many existing solutions to this issue, such as \citet{varselect}, involve explicitly or implicitly a Group LASSO \citep{group} calculation to produce this grouping effect. Incorporating this into LISO is possible, though it may produce a greatly increased computational burden. Instead, we shall apply ideas from \citet{adaptiveLASSO}. 

Consider the following two stage procedure -- we first conduct an ordinary LISO optimisation, arriving at an initial fit $\left(\seq{f^0_1}{f^0_p}\right)$. Then, we conduct a second LISO procedure, this time introducing covariate weights $\seq{w_1}{w_p}$ based on the first fit, and use the results of this as the output. We define the Adaptive LISO as the implementation of this, with $w_k = 1/\Delta{f^0_k}$, for $k = \seq{1}{p}$.

\begin{algorithm}
  \caption{Adaptive LISO}
  \label{alg:LISOadapt}
  \begin{algorithmic}[1]
    \STATE Calculate initial fit $f^0$ using LISO. (For instance, using Algorithm~\ref{alg:LISOback}.)
    \STATE Set $w_k = 1/\Delta(f^0_k)$, for $k = \seq{1}{p}$.
    \STATE Calculate, using e.g. Algorithm~\ref{alg:LISOback},
\[ \arg\min_{f_1, \ldots, f_p} \frac{1}{2}\norm{Y - \sum_{k=1}^p f_k(X^{(k)})}^2 +  \sum_{k=1}^p w_k\Delta(f_k), \textrm{ with } f_k  \in \mathcal{F}_{k}, \, k=1,\ldots,p. \]
    \STATE If necessary, set $f^0 = f$, and repeat from Step 2.
  \end{algorithmic}
\end{algorithm}

The analogy to the adaptive LASSO is that we apply a relaxation of the shrinkage for covariates with large fits in the initial calculation, and strengthen the shrinkage for covariates with small fits -- indeed, omitting entirely from consideration covariates initially fitted as zero. Usually, more than one reweighted calculation is not required.

The Adaptive LISO encourages grouping of the underlying LASSO optimisation because large steps contribute to relaxation of other steps in the same covariate. In addition, it means that we in general require less regularisation of true fits in order to shrink irrelevant covariates to zero, through the concavity of the implied overall optimisation, to which we are essentially calculating a Local Linear Approximation \citep{lla}. We will also always enhance sparsity through this procedure -- indeed, the fact that we reject straight away previously zero variables ensures the computational complexity of the method is usually at most equal to that of repeating the original LISO procedure for each iteration. 

It is, however, not clear what would be the best way to choose the tuning parameter introduced with each iteration of the process. We note that the discussants to \citet{lla} have recommended a scheme based on individual prediction error minimising cross validation at every step, and our empirical studies suggest that this can pose significant improvements over the basic LISO. In our experiments, we also implement a variant of the adaptive procedure, LISO-SCAD, where instead the weights are calculated with an implied group-wise SCAD penalty. LISO-SCAD and LISO-Adaptive hence both fit under a broad group of possible LISO-LLA procedures.

\subsection{Sign discovery and total variation penalty}\label{signsdisc}

Conventional isotonic regression focuses on the scenario where the monotonicity of the model function component in each covariate is known. However, this is not always realistic. Especially with large $p$, it may be the case that while we believe that the covariates contribute mostly in a monotonic way, we do not know, for at least some covariates, whether the covariate's component fit should be increasing, decreasing, or indeed even contribute non-monotonically. It is then perhaps reasonable to attempt to test for or estimate this monotonicity. This subject is dealt with by \citet{testingmono}, amongst others, with a focus on the univariate case. 

In higher dimensions, the problem is more difficult. One possible heuristic is to choose signs by a preliminary correlation check with the response. However, correlation is not invariant under general monotonic transformations, and examples exist where covariates have positive marginal effects, but, due to correlations between the covariates, turn out to have negative contributions in the final model.
%todo
Now, with the LISO, it is trivial to use the same LISO-backfitting method for calculation with relaxation, or selective relaxation of the monotonicity condition. In this case, the relaxed form is just minimising the residual sum of squares, penalised by the total variation of the fitted step function. In other words, we find the minimiser, with $\seq{f_1}{f_p}$ being univariate functions that have empirical mean zero and follow the specified combination of monotonicity constraints, of
\begin{equation}
 L_\lambda \left( (f_k)_{k=1}^p\right) := \frac{1}{2} \norm{Y - \sum_{k=1}^p f_k(X^{(k)})}^2 + \lambda \sum_{k=1}^{p} \Delta(f_k), \label{totalop}
\end{equation}
where $\Delta(f_k)$ is calculated using
\begin{equation*}
\Delta(f_k) = \begin{cases} 
f_k\left(\max X^{(k)}\right) - f_k\left(\min X^{(k)}\right) & \textrm{if $f_k$ is monotonically increasing,} \\
f_k\left(\min X^{(k)}\right) - f_k\left(\max X^{(k)}\right) & \textrm{if $f_k$ is monotonically decreasing,} \\
\sum_{i=2}^n \left \vert f_k\left(X^{(k)}_{(i)}\right) - f_k\left(X^{(k)}_{(i-1)}\right)\right\vert & \textrm{otherwise.}
\end{cases}
\end{equation*}

One way to implement this is to include reversed versions of non-monotonic covariates in the calculation, (and hence fitting a monotonically decreasing function to them as well as a monotonically increasing function) and then combine the fits with their corresponding twins after the calculation is complete.

\begin{definition}\label{decompdef}
 Let $(f_k)_{k=1}^p$ be any set of right continuous step functions with knots in the $k$-th covariate at $x^{(k)}_1 < \ldots < x^{(k)}_{n_k}$, and for each $k$, mean zero when evaluated at these knots. For $k = \seq{1}{p}$, define $f_k^+$ inductively as the right continuous step function with knot values
\begin{align*}
f_k^+\left(x^{(k)}_1\right) &= C^{(k)}_1 ,\\
\intertext{and for $j = \seq{1}{n_k -1}$,} 
f_k^+\left(x^{(k)}_{j+1 }\right) &= \begin{cases}
		  f_k^+\left(x^{(k)}_j\right) + f_k\left(x^{(k)}_{j+1}\right) - f_k\left(x^{(k)}_j\right) & \textrm{if $f_k\left(x^{(k)}_{j+1}\right) > f_k\left(x^{(k)}_j\right)$} \\
                  f_k^+\left(x^{(k)}_j\right) & \textrm{otherwise,}
                 \end{cases}
\end{align*}
with $C^{(k)}_1$ chosen so that $\sum_{i=1}^n f_k^+\left(x^{(k)}_i\right) = 0$.

Similarly, define $f_k^-$ as
\begin{align*}
f_k^-\left(x^{(k)}_1\right) &= C^{(k)}_2 ,\\
\intertext{and for $j = \seq{1}{n_k -1}$,} 
f_k^-\left(x^{(k)}_{j+1 }\right) &= \begin{cases}
		  f_k^-\left(x^{(k)}_j\right) + f_k\left(x^{(k)}_{j+1}\right) - f_k\left(x^{(k)}_j\right) & \textrm{if $f_k\left(x^{(k)}_{j+1}\right) < f_k\left(x^{(k)}_j\right)$} \\
                  f_k^-\left(x^{(k)}_j\right) & \textrm{otherwise,}
                 \end{cases}
\end{align*}
with $C^{(k)}_2$ chosen so that $\sum_{i=1}^n f_k^-\left(x^{(k)}_i\right) = 0$.
\end{definition}

Then, in the fully non-monotonic case, we have the following theorem:

\begin{theorem}\label{decomp}
Let  $(\widehat{g}_k, \widehat{h}_k)_{k=1}^p$ be the minimiser, with $g_k \in \mathcal{F}_k, h_k \in -\mathcal{F}_k, k = \seq{1}{p}$ of
\begin{equation}
 M_\lambda \left( (g_k, h_k)_{k=1}^p \right) := \frac{1}{2} \norm{Y - \sum_{k=1}^p \left(g_k\left(X^{(k)}\right) + h_k \left(X^{(k)}\right)\right)}^2 + \lambda \sum_{k=1}^{p}( \Delta(g_k) + \Delta(h_k)), \label{decompop}
\end{equation}

Then there is a one-to-one correspondence between such minimisers and minimisers $(\widehat{f}_k)$ to \eqref{totalop}. This correspondence is given by the decomposition above, so that $\widehat{g}_k = \widehat{f}^+_k, \widehat{h}_k = \widehat{f}^-_k$, and $\widehat{g}_k + \widehat{h}_k = \widehat{f}_k$, for all $k$.
\end{theorem}

An alternative implementation to using the above can be found by replacing the PAVA thresholding step in Algorithm 1 with a local thresholding style algorithm \citep{localadapt}. This can be slower, however, due to the computational burden involved with dealing with a covariate that would be taken exactly to zero, compared to checking \eqref{ifzero2} in the former case.

Now, extending Theorem~\ref{decomp}, the Adaptive LISO can provide an alternative way of dealing with the problem of sign discovery. Starting with an initial non-monotonic LISO fit, $\widetilde{f}_k$, $k = \seq{1}{p}$ say, we can conduct a second non-monotonic LISO fit, with covariate weights as in the Adaptive LISO case -- except that we treat the positive and negative component fits separately, with respect to the weights used.

Let $\widetilde{f}_k^+, \widetilde{f}_k^-$, $k = \seq{1}{p}$ be the decomposed version of the initial fit. The $M_\lambda$ approach will give us this decomposition directly, while we can apply the decomposition procedure from Definition~\ref{decompdef} to obtain the appropriate decomposition with the second implementation, or indeed an initial fit found by any other method. Then, setting $w^+_k = 1/\Delta(\widetilde{f}^+_k)$, $w^-_k = 1/\Delta(\widetilde{f}^-_k)$ we find the LISO adaptive sign discovery solution to be simply $\widehat{g} + \widehat{h}$, where $\widehat{g}_k, -\widehat{h}_k \in \mathcal{F}_k\, k= \seq{1}{p}$ are solutions of

\[
\arg \min_{\substack{\seq{g_1}{g_p} \\ \seq{h_1}{h_p}}} \frac{1}{2} \norm{Y - \sum_{k=1}^p\left( g_k(X^{(k)}) + h_k(X^{(k)})\right)}^2 + \lambda \sum_{k=1}^{p} (w^+_k \Delta(g_k) + w^-_k\Delta(h_k)).
\]

Thus, as well as the effect seen in the adaptive LISO, where we have strengthened shrinkage of small function fits towards zero, functions with small negative or positive components in the initial fit will be shrunk towards an monotonically increasing or decreasing function respectively. 

\section{Numerical results}\label{simusec}

We will present a series of numerical examples designed to illustrate the effectiveness of the LISO in handling additive isotone problems. The experiments are calculated in R, using a standard desktop workstation. The full path solutions are found using a LISO modification to the Lars algorithm \citep{efronlars}, while the larger comparison studies and fits are conducted using an implementation of the backfitting algorithm, with a logarithmic grid for the tuning parameter.

\subsection{Example LISO fits}

The following examples, conducted on single datasets, illustrate the performance of the algorithm.
\subsubsection{Boston Housing dataset}

The Boston Housing dataset, as detailed in \citet{bostonhousing}, is a dataset often used in the literature to test estimators -- see e.g. \citet{elementsofsl}. The dataset comprises of $n=506$ observations of 13 covariates, plus one response variable, which is the median house prices at each observation location. The response is known to be censored at the value 50, while the covariates range from crime statistics to discrete variables like index of accessibility to highways. We use here the version included in the R \texttt{MASS} library, though we shall discard the indicator covariate \texttt{chas}, for ease of presentation. (Experiments suggest that this variable does not have a great effect on the response, in any case.) 

As suggested in \citet{ravispam}, we will test the selection accuracy of the model by adding $U(0,1)$ irrelevant variables. We add 28, so that our final $p=40$. Since signs are not known, we will apply the sign discovery version of the LISO from Section~\ref{signsdisc}, by first conducting a non-monotonic total variation fit, and then a weighted second fit. Tuning parameters are chosen by two 10-fold cross validations.

Our selected model, finally, is
\begin{align*}
Y =& \alpha + f_1(\mathtt{crim})+ f_2(\mathtt{nox})+ f_3(\mathtt{rm}) + f_4(\mathtt{dis})+ f_{5}(\mathtt{tax})+f_{6}(\mathtt{ptratio})+ f_{7}(\mathtt{lstat}) + \varepsilon.
\end{align*}

\begin{figure}[ht]
 \centering
 \includegraphics[scale = 0.5]{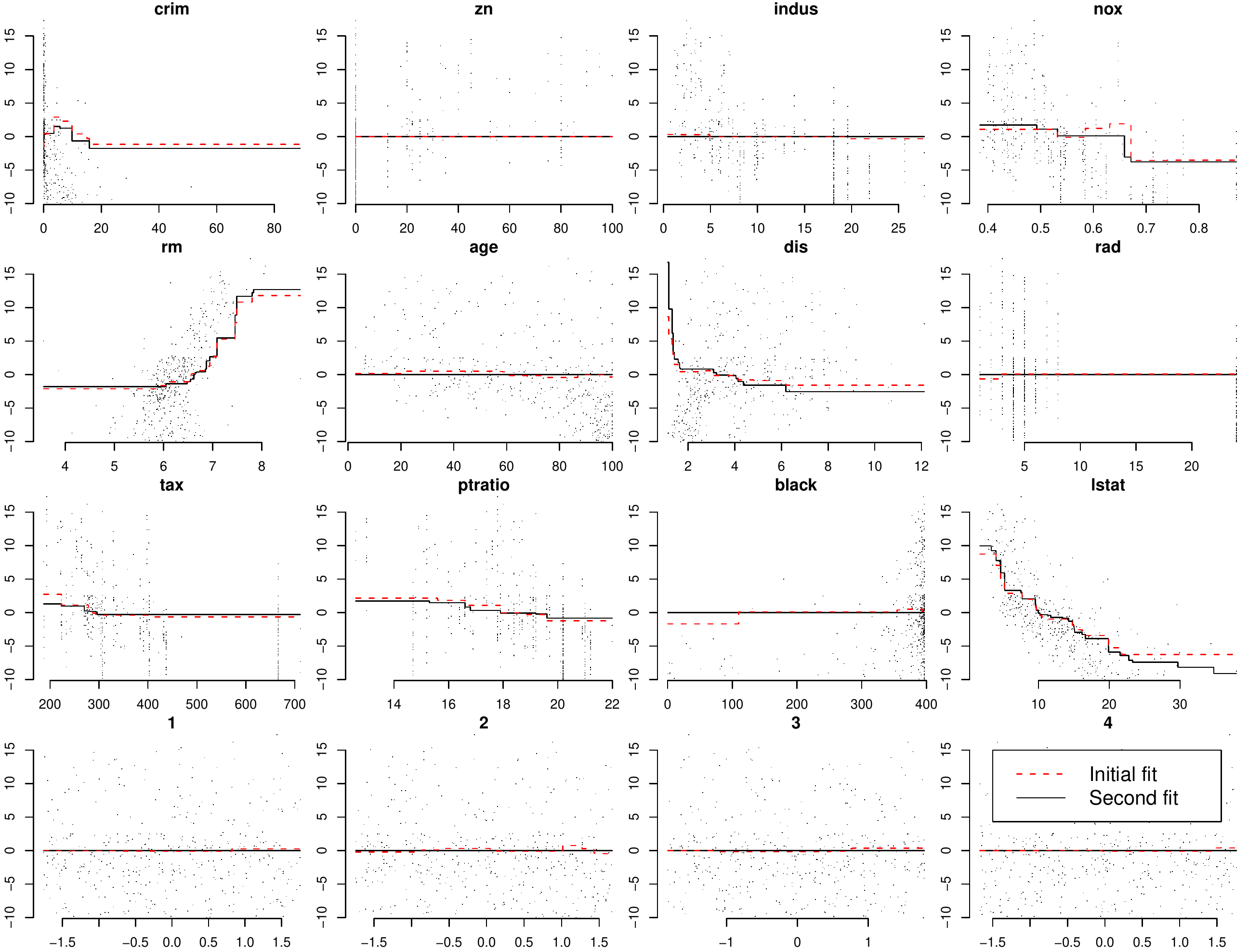}
 % threshlow.eps: -1x-1 pixel, 300dpi, -0.01x-0.01 cm, bb=0 0 815 942
 \caption{Fitted component functions on the Boston Housing dataset, for covariates originally present in the data plus four others. The dashed line shows the selected model after the first LISO step, while the solid black line shows the final result of the adaptive sign finding procedure. The single step fit produced additional non-zero fits in some of the artificial covariates, which are not shown, while the two step procedure fit all of them as zero.}
 \label{fig:bostonfit}
\end{figure}

The remaining covariates are judged to have an insignificant effect on the response, with zero regression fits. $f_3$ was found to be monotonically increasing, $f_1$ non-monotonic, and the remaining functions monotonically decreasing. The full results are shown in Figure~\ref{fig:bostonfit}.

We see in our experiments that for higher values of $\lambda$, we successfully remove all the irrelevant variables, and end up with only a small number of selected variables to explain the response. However, in the one step procedure, the amount of shrinkage required is often large. With cross validation as a criterion, we do choose a $\lambda$ that involves some irrelevant variables as well, though these are in general small in magnitude. A second step greatly improves the model selection characteristics, as well as creating monotonicity which is often absent in the first step.

It is interesting to contrast our fit with the findings from using SpAM \citep{ravispam}. Bearing in mind that our problem was in some sense more difficult, since we had 12 original covariates instead of 10 (\texttt{rad} and \texttt{zn} were not included in the SpAM study), and 28 artificial covariates instead of 20, our findings are largely similar. In addition to the covariates selected in SpAM, we add a fairly large effect from \texttt{nox}, and smaller effects in \texttt{dis} and \texttt{tax}. The most significant fits on \texttt{rm} and \texttt{lstat} are very similar, though the LISO fit is clearly less smooth. However, while almost all of the fits from SpAM exhibit non-monotonicity, the LISO fit we have found is mostly monotone, aside from the fit in \texttt{crim}.

The non-monotonicity found in \texttt{crim} may seem problematic, given the interpretation of that covariate as a crime rate. While, nevertheless, this is a characteristic present in the conditional residuals, perhaps it would be reasonable to impose a monotonicity constraint instead.

\subsubsection{Artificial dataset}

We are also interested in the success of LISO in correctly selecting variables for varying levels of $n$ and $p$. We adopt the following setup -- we generate pairs $X \in \R^{n \times p}, Y \in \R^n$ by
\begin{align*}
 X_{ij} &\sim \mathit{Uniform}\left(0,1\right) \\
 Y_{i} &= 2 \left(X^{(1)}_i\right)_+ ^2 + X^{(2)}_{i} + \sign\left(X^{(3)}\right)  \left|X^{(3)}_i\right|^{1/5} + 2 I_{\lbrace X^{(4)}_i > 0\rbrace} + \varepsilon_i
\end{align*}
with $n = 1024,$ $p = 1024$, independent $\varepsilon_i \sim N(0,1)$. The covariates are then centred and standardised to have mean zero and variance 1, and Y is centred to have mean zero.

For $p' = 32, 64, 128, 256, 512, 1024$, $n' = 5, 10, 15, \ldots$ , we then take as $X', Y'$ subsets of $X, Y$ corresponding to the first $p'$ columns of $X$, and random samples without replacement of $n'$ rows of $X,Y$. Hence we consider the problem of correctly finding $4$ true variables, from amongst $p'$ potential ones, based on $n'$ observations. We quantify the success of LISO by looking at the proportion of $50$ replications where the algorithm, for at least one value of $\lambda$, produces an estimate where the true covariates have at least one step while the other covariates are taken to zero. (We adopt this framework so as to reduce the additional noise from generating a complete new random dataset with each attempt.)

\begin{figure}[ht]
\vspace{-15pt}
 \centering
 \includegraphics[scale = 0.5]{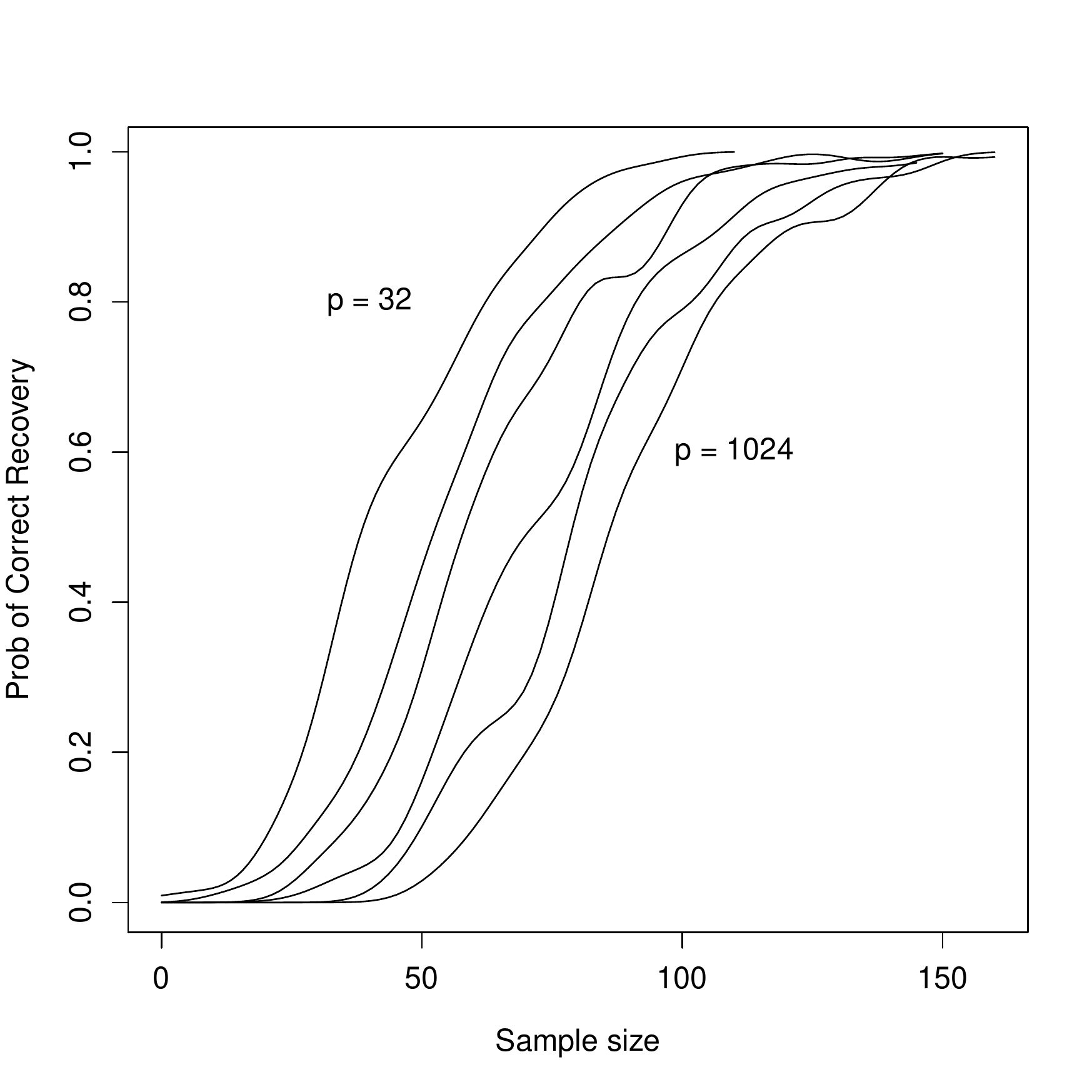}
 % threshlow.eps: -1x-1 pixel, 300dpi, -0.01x-0.01 cm, bb=0 0 815 942
\vspace{-15pt}
 \caption{Probabilities of correct sparsity recovery with $4$ true nonlinear but monotonic covariates, $SNR = 4$. Each line shows how the recovery probability changes as the sample size $n$ changes for a single value of $p$, taking values $2^5, \ldots, 2^{10}$.}
 \label{fig:ntest}
\end{figure}

Figure~\ref{fig:ntest} gives these results. As we can see, as in a variety of LASSO-type algorithms \citep{wainwrightsharp}, there is a sharp threshold between success and failure in recovery of sparsity patterns as a function of $n$. Moreover, as we increase $p$ exponentially, the required number of observations $n$ increases much more slowly, thus implying that $p \gg n$ recovery is possible. 

\begin{figure}[ht]
 \centering
 \includegraphics[scale = 0.5]{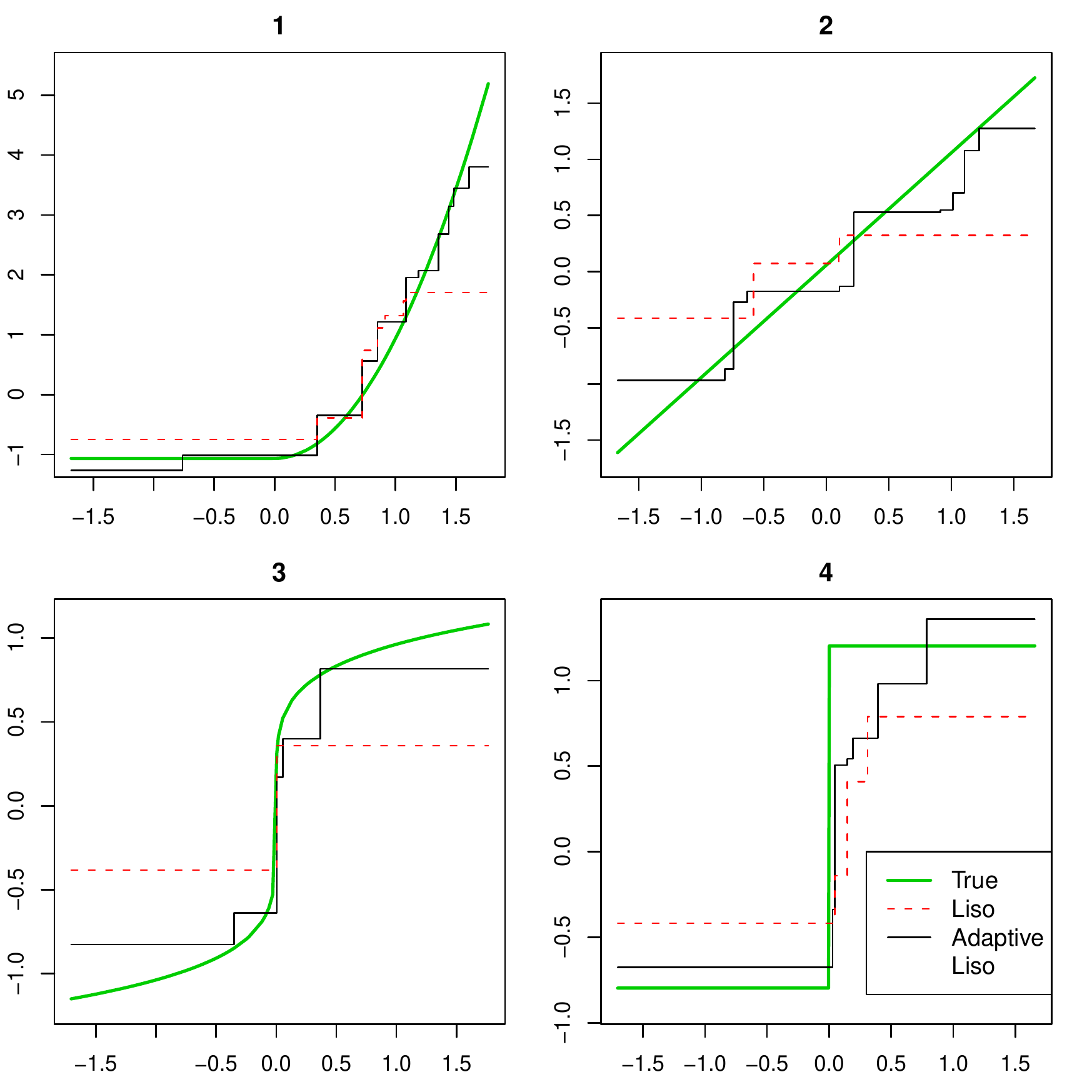}
 % threshlow.eps: -1x-1 pixel, 300dpi, -0.01x-0.01 cm, bb=0 0 815 942
 \caption{Example LISO covariate fits, for $n = 180$, $p=1024$. The true component functions are given by the thick line, while the dashed line gives the raw LISO fit for the smallest amount of regularisation required to bring spurious fits in irrelevant covariates to zero. The solid black line shows a fit made by the adaptive liso, using tuning parameters found by cross validation. The fitted and true model functions for all 1020 remaining covariates are all constant zero.}
 \label{fig:linfitexample}
\end{figure}

Figure~\ref{fig:linfitexample} gives an example of LISO fits arising from this simulation. The dashed lines shows the results of the LISO under the minimum regularisation required for correct sparsity recovery -- note the high level of shrinkage required to shrink the other variables to zero. This shrinkage exhibits itself as not only a thresholding on the ends of the component fits, which we have seen in the univariate case, a , but also an additional loss of complexity in the middle parts of each component fit. We can avoid these shrinkages by using this initial result to perform the Adaptive LISO, in the solid black line, thus greatly improving the fit while still keeping the correct sparsity pattern recovery. As an added bonus, we get good results here with the Adaptive LISO even without using knowledge of the true process that generated the data.%Regardless, we would expect that due to the thresholding effect, the LISO would perform best on predicting new observations in the middle of covariate ranges.

\subsection{Comparison studies}

We shall now compare LISO to a range of other procedures in some varying contexts. Varying $f$ between scenarios, consider generating pairs $X, Y$ by, for each repetition,

\begin{align*}
X_i^{(j)} &\sim \mathit{Uniform}\left(-1, 1\right), \quad i = 1, \ldots, n,\, j = 1,\ldots, p \\
\varepsilon_i &\sim N(0, 1), \quad i = 1, \ldots, n\\
Y_i &= f(X_i) + \sigma \varepsilon_i, \quad i = 1, \ldots, n.
\end{align*}

100 repetitions were done of each combination of model and noise level, with $\sigma$ chosen to give $SNR = 1, 3$ or $7$, plus one further case where we have $SNR = 3$ but $X$ is instead generated to have stronger correlation between the covariates, as a rescaled (to the range $(-1,1)$) version of $\Phi(Z), Z \sim N(0, \Sigma)$, with $\Sigma_{ij} = 2^{-|i-j|}$.

For comparison, we will compare the performance of LISO and LISO-LLA (both Adaptive and SCAD), calculated using the backfitting algorithm, to
\begin{itemize}
 \item Random Forests (RF), from \citet{randf}. A tree based method using aggregation of trees generated using a large number of resamplings.
\item Multiple Adaptive Regression Splines (MARS), from \citet{mars:1991}, using the \texttt{earth} implementation in R. A method using greedy forward/backward selection with a hockeystick shaped basis. We use a version restricted to additive model fitting.
 \item Sparse Additive Models (SpAM), from \citet{ravispam}. A similar group LASSO based method using soft thresholding of component smoother fits.
 \item Sparsity Smoothness Penalty (SSP), from \citet{lukasham}. A group LASSO based method using two penalties -- a sparsity penalty and an explicit smoothness penalty.

\end{itemize}

For the choice of tuning parameter in all algorithms, we take the value that minimises the prediction error on a separate validation set of the same size as the training set. (Note that in the case of SSP, due to the slowness of finding two separate tuning parameters, we instead perform a small number of initial full validation runs for each scenario. We then plug in the averaged smoothness tuning parameter in all following runs, optimising for only the sparsity parameter.)

We record both the mean value across runs of the MSE on predicting a new test set (generated without noise), and, in brackets, the mean relative MSE, defined for the $k$-th algorithm on each individual run as
\[MSE^k_{\mathbf{Relative}} := \frac{MSE^k}{\min_{j = 1, \ldots,7}MSE^j}. \]

\subsubsection{All components linear}
In this case, we have the response being just a scaled sum of $k=5$ randomly chosen covariates, plus a noise term. $n = 200$, $p = 50$ overall. In the test set, the variance of the response (and hence the MSE of a constant prediction) was approximately 1.7.

\begin{center}
\begin{tabular}{l|l|l|l|l}
\textbf{Algorithm} & \textbf{SNR = 7} & \textbf{SNR = 3} & \textbf{SNR = 1} & \textbf{SNR = 3, Correlated} \\
\hline

LISO &0.113 (4.70) &0.186 (3.33) &0.358 (2.41) &0.203 (3.43) \\
LISO-Adaptive &0.070 (2.94) &0.118 (2.18) &0.242 (1.62) &0.134 (2.27) \\
LISO-SCAD &0.113 (4.71) &0.186 (3.33) &0.437 (3.00) &0.202 (3.41) \\
SpAM &0.082 (3.29) &0.149 (2.57) &0.346 (2.24) &0.159 (2.59) \\
SSP &0.026 (1.00) &0.061 (1.00) &0.167 (1.02) &0.065 (1.00) \\
RF &0.286  (11.97) &0.319  (5.85) &0.504  (3.36) &0.361 (6.21) \\
MARS &0.146 (6.28) &0.354 (6.72) &1.027 (6.91) &0.417 (7.19) \\

\end{tabular}
\end{center}

Because of the sparsity and additivity in the data, all LASSO-like methods do better than RF, a pattern that continues in all of these simulation studies. Indeed, due to the random selection of covariates in the RF algorithm, the presence of spurious covariates seems to produce a phenomenon of excess shrinkage, which can be clearly see in plots of fitted values versus response values. Using the scaling corrections provided in the R implementation improves things, but not to a great extent. MARS, similarly, has difficulty in finding the correct variables. With such large $p$, the set of possible hockey stick bases MARS has to search through is very large, and hence the underlying greedy stepwise selection component of the algorithm is in general unsuccessful at handling this problem.

Amongst the LASSO-like methods, perhaps unsurprisingly, the SSP method performs by far the best, owing to the large degree of smoothness in the true model function. LISO-Adaptive is second best, however, beating SpAM even though it does not have an internal smoothing effect. The basic LISO method itself underperforms, perhaps because it does not strongly enforce sparsity amongst the original covariates.

Unexpectedly, LISO-SCAD performs fairly equivalently to the LISO itself in this and all following simulations. A likely explanation is that for sufficient regularisation to take place to take spurious covariates to zero, the penalty function is such that the solution lies mostly on the part of the penalty where it is identical to the original total variation penalty.

The introduction of a moderate amount of correlation does not greatly affect the performance of any of the algorithms.

\subsubsection{Mixed powers}
In this case, the response has a more complex relation to the covariates:

\begin{align*}
 Y_i &= \sum_{k =1}^5 f_k(X^{(a_k)}_i) + \sigma \varepsilon_i \\
 f_1(x) &= \sign(x + C_1) \left\vert x+C_1 \right\vert^{0.2} \\
f_2(x) &= \sign(x + C_2) \left\vert x+C_2 \right\vert^{0.3} \\
f_3(x) &= \sign(x + C_3) \left\vert x+C_3 \right\vert^{0.4} \\
f_4(x) &= \sign(x + C_4) \left\vert x+C_4 \right\vert^{0.8} \\
f_5(x) &= x + C_5 \\
\end{align*}

In this case, we have again $n=200, p=50$. $\seq{C_1}{C_5}$ are small shifts, randomly generated as $\mathit{Uniform}(-1/4, 1/4)$, and $\seq{a_1}{a_5}$ are covariates randomly chosen without replacement. In the test set, the variance of the response was approximately 2.6.

\begin{center}
\begin{tabular}{l|l|l|l|l}
\textbf{Algorithm} & \textbf{SNR = 7} & \textbf{SNR = 3} & \textbf{SNR = 1} & \textbf{SNR = 3, Correlated} \\
\hline
LISO &0.128 (1.49) &0.230 (1.50) &0.459 (1.41) &0.255 (1.50) \\
LISO-Adaptive &0.088 (1.01) &0.160 (1.00) &0.352 (1.06) &0.177 (1.01) \\
LISO-SCAD &0.128 (1.49) &0.229 (1.49) &0.587 (1.82) &0.254 (1.50) \\
SpAM &0.157 (1.83) &0.267 (1.75) &0.539 (1.68) &0.285 (1.69) \\
SSP &0.126 (1.47) &0.226 (1.49) &0.429 (1.33) &0.252 (1.51) \\
RF &0.358 (4.21) &0.450 (2.96) &0.721 (2.26) &0.495 (2.96) \\
MARS &0.319 (3.78) &0.678 (4.54) &1.936 (6.32) &0.783 (4.71) \\

\end{tabular}
\end{center}

With the new, non-linear model function, the LISO and LISO-SCAD now perform equally as well as the SSP, while the adaptive LISO performs significantly better, being the best in almost all runs. All four methods outperform SpAM, and greatly outperform RF and MARS.

 In this case, the explanation is that for fractional powers, the component functions are relatively flat in the extremes of the covariate range, with most of the variation occuring in the middle of the range. SpAM and SSP are unable to capture the sharp transition point of the small root functions without introducing inappropriate variability at the ends of the fit, and hence both perform significantly worse than previously. The LISO based methods, however, do not explicitly smooth the fit and only threshold the extremes. Being thus adapted to this sort of function, they actually improve their performance in proportional terms relative to the variance of the test set.

\subsubsection{Mixed powers, large p}
In this scenario, our model is the same as before, save that we have many more spurious covariates, resulting in $n=200, p= 200$. The variance of the test response is unchanged at approximately 2.6.

\begin{center}
\begin{tabular}{l|l|l|l|l}
\textbf{Algorithm} & \textbf{SNR = 7} & \textbf{SNR = 3} & \textbf{SNR = 1} & \textbf{SNR = 3, Correlated} \\
\hline
LISO &0.166 (1.89) &0.283 (1.86) &0.638 (1.78) &0.286 (1.84) \\
LISO-Adaptive &0.090 (1.00) &0.156 (1.00) &0.384 (1.01) &0.160 (1.01) \\
LISO-SCAD &0.169 (1.93) &0.292 (1.91) &0.935 (2.71) &0.296 (1.90) \\
SpAM &0.201 (2.32) &0.329 (2.17) &0.779 (2.21) &0.331 (2.14) \\
SSP &0.156 (1.78) &0.274 (1.80) &0.604 (1.73) &0.274 (1.78) \\
RF &0.504 (5.86) &0.588 (3.86) &0.992 (2.84) &0.593 (3.84) \\
MARS &0.805  (9.27) &1.704  (11.49) &4.707  (13.84) &1.763 (11.60) \\
\end{tabular}
\end{center}

In this case, LISO preserves its superiority. Due to the effect of high dimensionality, all algorithms see their performance decline - except the adaptive LISO, which has an increased MSE of less than 3\% in the low noise case. This is due to the adaptive step, which retains a very sparse fit, picking the relevant variables even as the number of predictors grows.

\section{Discussion}\label{concsec}

We have presented here a method of extending ideas from LASSO on linear models to the framework of non-parametric estimation of isotonic functions. We have found that in many contexts, it inherits the behaviour of the LASSO in that it allows sparse estimation in high dimensions. By using our backfitting procedure, we have also shown empirically that it can be very competitive with many current methods, both in terms of computational time and memory requirements, and in terms of predictive accuracy. The precise criteria that govern its success would require further work, and it would be interesting to see if similar LASSO-style oracle results apply.

In addition, we find that an LLA/adaptive scheme is highly effective and efficient at improving the algorithm in a two step approach, producing sparser results and very high predictive accuracy. Further adaptations allow the LISO method to be used when monotonicity is assumed but the direction of the monotonicity is not know. To the authors' knowledge, this has not been attempted previously in this type of problem, and it would be interesting to see if LLA and similar concave penalty procedures can produce effective replacements for the group LASSO in the underlying calculation of non-parametric LASSO generalisations.

\appendix
\section{Appendix: Proofs of theorems}
\subsection*{Proof of Theorem~\ref{thresh1}}
Our methodology is to show that adding boundary constraints to constrained or unconstrained isotonic regression problems result in unique solutions that are simply Winsorised PAVA estimates, and then demonstrate a method of constructing any LISO solution, in the univariate case, though boundary constraints. We prove first the following lemma, which provides an induction step in our eventual argument:
\begin{lemma}\label{bottomad}
Suppose for $A \leq B$, $X = (X_1, \ldots, X_n), Y = (Y_1,\ldots,Y_n)$, $X_i \in \R$ for all $i$, the Winsorized PAVA,
\[f_1(x) = \haf{>A,<B}(x) := \begin{cases}
A & \textrm{if }\haf{\mathit{PAVA}}(x) < A\\
B & \textrm{if }\haf{\mathit{PAVA}}(x) > B\\
\widehat{f}_{\mathit{PAVA}}(x) & \mathrm{otherwise}.
\end{cases}\]
solves the boundary constrained isotonic regression problem,
\begin{equation}
\min_{f} \norm{Y-f(X)}^2 \quad \textrm{such that $f$ monotone, $A \leq f(x) \leq B, \quad \forall x$.} \label{A1op} 
\end{equation}
Then for $A \leq A' \leq B' \leq B$, $f_2 \equiv \haf{>A',<B'}$ solves the further constrained isotonic regression problem
\begin{equation}
\min_{f} \norm{Y-f(X)}^2 \quad \textrm{such that $f$ monotone, $A' \leq f(x) \leq B', \quad \forall x$.} \label{A2op} 
\end{equation}

Further, this solution is unique, in terms of its fitted values $f(X_1), \ldots, f(X_n)$.
\end{lemma}
\begin{proof}[Proof of Lemma~\ref{bottomad}]
It suffices to prove the lemma for the case of $A \leq A' \leq B' = B$, since the argument for $A = A' \leq B' \leq B$ is identical, and we can proceed to the full Lemma by adding the top and bottom constraints one by one.

Note that, specifying $f$ through the fitted values $f(X_1), \ldots, f(X_n)$, $\norm{Y-f(X)}^2$ is strictly convex when considered as a function of $f$, and the constraints give a convex feasible set. Hence, for any combination of $A,B$, solutions must exist and be unique at $X_1, \ldots, X_n$. 

Therefore, let $g$ be the solution of the optimisation~\eqref{A2op}, for $A \leq A' \leq B' = B$. Suppose for contradiction that $g \not\equiv f_2 \equiv \haf{>A',<B'}$.

Let $u_f, u_g$ be the points where the functions $f_2, g$ respectively exceed $A'$.
\begin{align*}
u_f &= \inf \left\lbrace X_i\, \textrm{ s.t. }\, f_2(X_i) > A' \right\rbrace \\
u_g &= \inf \left\lbrace X_i\, \textrm{ s.t. }\, g(X_i) > A' \right\rbrace. 
\end{align*}

Then we have two cases.

(a) If $u_f \leq u_g$, then consider a new function $\widetilde{f}$ where
\begin{equation}
\widetilde{f}(x) = \begin{cases} 
f_1(x) & \textrm{if} \quad x < u_f  \\
g(x) & \textrm{if} \quad x \geq u_f. 
\end{cases}
\end{equation}

$\widetilde{f}$ would be an increasing function satisfying the conditions of \eqref{A1op}. Because $g \not\equiv f_2$, and $g$ and $f_2$ are both equal to $A'$ when restricted to $\{x : x < u_f\}$, it must be the case that $g \not\equiv f_2 \equiv f_1$ when restricted to $\{x : x \geq u_f\}$.  Therefore, $\widetilde{f} \not\equiv f_1$. The residual sum of squares is then, applying \eqref{A2op} optimality of $g$,
\begin{align*}
\norm{Y - \widetilde{f}(X)}^2 &= \sum_{X_i<u_f} \left(Y_i- f_1(X_i)\right)^2 +  \sum_{X_i\geq u_f}\left(Y_i- g(X_i)\right)^2  \\
&=\sum_{X_i<u_f} \left(Y_i- f_1(X_i)\right)^2 +  \norm{Y- g(X)}^2 - \sum_{X_i< u_f}\left(Y_i-A'\right)^2  \\
&\leq \sum_{X_i<u_f} \left(Y_i- f_1(X_i)\right)^2 +  \norm{Y- f_2(X)}^2 - \sum_{X_i< u_f}\left(Y_i-A'\right)^2  \\
&= \sum_{X_i<u_f} \left(Y_i- f_1(X_i)\right)^2 +  \sum_{X_i\geq u_f}\left(Y_i- f_1(X_i)\right)^2 \\
&= \norm{Y - f_1(X)}^2.
\end{align*}

Therefore, $\widetilde{f}$ is optimal for \eqref{A1op}. This contradicts uniqueness and \eqref{A1op} optimality of $f$.

(b) If $u_f > u_g$, then if we define $\widetilde{f}$ this time as
\begin{equation}
\widetilde{f}(x) = \begin{cases} 
f_1(x) & \textrm{if}\quad  x < u_g  \\
g(x) & \textrm{if}\quad  x \geq u_g, 
\end{cases}
\end{equation}
we obtain another increasing function satisfying the conditions of \eqref{A1op}. As before, because we have assumed that $g \not\equiv f_2$, and yet $g \equiv f_2 \equiv A'$ for $\{x : x< u_g\}$,  $g \not\equiv f_2 \equiv f_1$ when restricted to $\{x : x\geq u_g\}$. This means that $\widetilde{f} \not\equiv f_1$, so unique optimality of $f_1$ versus $\widetilde{f}$ means that

\begin{align*}
\sum_{X_i\geq u_g}\left(Y_i- f_1(X_i)\right)^2  &= \norm{Y - f_1(X)}^2 - \sum_{X_i< u_g}\left(Y_i- f_1(X_i)\right)^2 \\
&< \norm{Y - \widetilde{f}(X)}^2 -  \sum_{X_i< u_g}\left(Y_i- f_1(X_i)\right)^2 \\
&= \sum_{X_i\geq u_g}\left(Y_i- g(X_i)\right)^2.
\end{align*}

In addition, because $u_f > u_g$, setting $\delta= (g(u_g) - A') / (g(u_g) - f_1(u_g)) $ makes $\widetilde{g}$, defined as
\begin{equation}
\widetilde{g}(x) = \begin{cases} 
A' & \textrm{if}\quad  x < u_g  \\
(1-\delta) g(x) + \delta f_1(x) & \textrm{if}\quad  x \geq u_g, 
\end{cases}
\end{equation}
an increasing function satisfying the conditions of \eqref{A2op}. By definition, $g(u_g) > A'$ and $f_1(u_g) < A'$, so $\delta \in (0,1)$, implying that $\widetilde{g}$ is a nontrivial convex combination of $g$ and $f_1$, when restricted to $\{x : x \geq u_g\}$.

But by convexity, 
\begin{align*}
 \norm{Y - \widetilde{g}(X)}^2 &= \sum_{X_i<u_g} \left(Y_i -A'\right)^2 +  \sum_{X_i \geq u_g} \left(Y_i- (1-\delta) g(X_i) - \delta f_1(X_i)\right)^2 \\
&< \sum_{X_i<u_g} \left(Y_i -A'\right)^2 +  \sum_{X_i \geq u_g} \left(Y_i- g(X_i)\right)^2 =\norm{Y - g(X)}^2.
\end{align*}

This contradicts optimality of $g$.

Therefore, $g \equiv f_2$ is the unique solution to \eqref{A2op}.
\end{proof}

\begin{proof}[Proof of Theorem~\ref{thresh1}]

We can prove Theorem~\ref{thresh1} as a simple corollary.

When $\lambda = 0$, our objective function $L_\lambda$ is strictly convex and quadratic, and indeed is the same as the PAVA optimisation. Hence, an unique optimal solution exists, and is given by $\haf{0} = \haf{\mathit{PAVA}}$. Set $A_0 = -\infty, B_0=\infty$. Then $\haf{0} \equiv \haf{>A_0, <B_0}$ is feasible for, and so, must also solve the constrained optimisation \eqref{A1op}, with constraints at infinity.

For $\lambda > 0$, $L_\lambda$ and the domain we maximise it in are both still convex, with $L_\lambda$ strictly convex and increasing away from the origin outside of a neighbourhood. Therefore, an unique bounded solution $\haf{\lambda}$ must exist. Set $A_\lambda, B_\lambda$ to be the upper and lower bounds of this solution.
\[
A_\lambda = \min_i \haf{\lambda}(X_i),\quad B_\lambda = \max_i \haf{\lambda}(X_i).
\]

Then consider the solution to the constrained isotonic least squares problem
\begin{equation}
\widetilde{f}_\lambda = \arg\min_f \norm{Y-f(X)}^2 \quad \textrm{such that $f$ monotone, $A_\lambda \leq f(x) \leq B_\lambda, \quad \forall x$.} \label{doublecons}
\end{equation}

$\Delta(\widetilde{f}_\lambda) \leq B_\lambda - A_\lambda = \Delta(\haf{\lambda})$, and since $\haf{\lambda}$ is feasible for \ref{doublecons}, $\norm{Y-\widetilde{f}_\lambda(X)}^2 \leq \norm{Y-\haf{\lambda}(X)}^2$.

Therefore, $\widetilde{f}_\lambda$ is optimal for \eqref{lambdaop}. Hence, by uniqueness, $\widetilde{f}_\lambda \equiv \haf{\lambda}$, so for suitable $A_\lambda, B_\lambda$ it suffices to solve the bound constrained least squares optimisation \eqref{doublecons}, to find the LISO solution.

But by Lemma~\ref{bottomad}, the solution of \eqref{doublecons} for $>A_\lambda <B_\lambda$ is just the Winsorised PAVA solution $\haf{>A_\lambda, <B_\lambda}$, so $\haf{\lambda} \equiv \widetilde{f}_\lambda \equiv \haf{>A_\lambda, <B_\lambda}$.
\end{proof}

\subsection*{Proof of Theorem~\ref{threshfind}}

\begin{proof}
For $\lambda = 0$, $\haf{0} \equiv \haf{\mathit{PAVA}}$, so choosing $A_0 = \min(\haf{\mathit{PAVA}}(x))$, $B_0 = \max(\haf{\mathit{PAVA}}(x))$ is clearly a optimum for \eqref{lambdaop} that satisfies \eqref{Bmini} and \eqref{Amini}. Assume therefore $\lambda > 0$.

Now, from \citet{underorder}, the unregularised PAVA solution $\haf{\mathit{PAVA}}$, considered as a right continuous step function, is an example of a regressogram. In other words, there exists a partition into disjoint intervals of $\R$, $\seq{P_1}{P_m}$, with the value of $\haf{\mathit{PAVA}}(x)$ on each interval being the mean of the observed $Y$ for $X$ falling within the interval.

\[\haf{\mathit{PAVA}}(x) = \frac{\sum_{i=1}^n Y_iI_{\{X_i \in P_j\}} }{\sum_{i=1}^n I_{\{X_i \in P_j\}}} = \haf{\mathit{PAVA}}(P_j) \quad\textrm{for all $x \in P_j$}\]

Using $\haf{0} \equiv \haf{\mathit{PAVA}}$, the LASSO criterion for the thresholded function $\haf{>A, <B}$ then can be written as,
\begin{align*}
L_\lambda(\haf{>A, <B}) &= \frac{1}{2}\norm{Y - \haf{>A, <B}}^2 + \lambda(B-A) \\
&= \frac{1}{2}\sum_{i=1}^n \left((Y_i - \haf{0}(X_i))^2 I_{\{\haf{0}(X_i) \in [A,B]\}} + (A - Y_i)^2I_{\{\haf{0}(X_i) < A\}} + (Y_i-B)^2I_{\{\haf{0}(X_i) > B\}}\right) \\
&\quad + \lambda(B-A) \\
&= \frac{1}{2}\norm{Y - \haf{0}}^2 + \frac{1}{2}\sum_{i=1}^n \left((A - \haf{0}(X_i))^2 + 2(A - \haf{0}(X_i))( \haf{0}(X_i) - Y_i) \right)I_{\{\haf{0}(X_i) < A\}} \\
&\quad+ \frac{1}{2}\sum_{i=1}^n \left((\haf{0}(X_i) -B)^2 + 2(\haf{0}(X_i) -B)(Y_i - \haf{0}(X_i))\right)I_{\{\haf{0}(X_i) > B\}} + \lambda(B-A) \\
&= \frac{1}{2}\norm{Y - \haf{0}}^2 + \frac{1}{2}\sum_{i=1}^n\left( (A - \haf{0}(X_i))_+^2+ (\haf{0}(X_i) -B)_+^2 \right) + \lambda(B-A) \\
&\quad+ \sum_{j=1}^m\sum_{i=1}^n  \left((\haf{0}(X_i) -B)(Y_i - \haf{0}(X_i))I_{\{\haf{0}(X_i) > B\}}\right)I_{\{X_i \in P_j\}} \\ 
&\quad+ \sum_{j=1}^m\sum_{i=1}^n  \left((A - \haf{0}(X_i))( \haf{0}(X_i) - Y_i)I_{\{\haf{0}(X_i) < A\}}\right)I_{\{X_i \in P_j\}} \\
&= \frac{1}{2}\norm{Y - \haf{0}}^2 + \frac{1}{2}\sum_{i=1}^n\left( (A - \haf{0}(X_i))_+^2+ (\haf{0}(X_i) -B)_+^2 \right) + \lambda(B-A) \\
&\quad+ \sum_{j=1}^m (\haf{0}(P_j) -B)_+\sum_{i=1}^n  \left((Y_i - \haf{0}(X_i))\right)I_{\{X_i \in P_j\}} \\ 
&\quad+ \sum_{j=1}^m (A - \haf{0}(P_j))_+\sum_{i=1}^n  \left(( \haf{0}(X_i) - Y_i)\right)I_{\{X_i \in P_j\}} \\
&= \frac{1}{2}\norm{Y - \haf{0}}^2 + \frac{1}{2}\sum_{i=1}^n\left( (A - \haf{0}(X_i))_+^2+ (\haf{0}(X_i) -B)_+^2 \right) + \lambda(B-A).
\end{align*}

We seek a minimum to this with $A \leq B$. Differentiating in $A$ and $B$, and setting equal to zero, gives,
\begin{align}
  \sum_{i = 1}^n  (\haf{0}(X_i) - B)_+ &= \lambda \label{Bmin} \\ 
  \sum_{i = 1}^n  (A-\haf{0}(X_i))_+ &= \lambda \label{Amin}. 
\end{align}

Note that the left hand side in both cases is a piecewise linear, continuous and monotone (indeed, decreasing in $B$ and increasing in $A$) function of the threshold, equalling zero for $A = A_0, B = B_0$. Further, since the PAVA is a regressogram, $\sum_{i = 1}^n  \haf{0}(X_i) = \sum_{i = 1}^n Y_i$, so
\[\sum_{i = 1}^n  (\haf{0}(X_i) - \overline{Y})_+ = \sum_{i = 1}^n  (\overline{Y} -\haf{0}(X_i))_+ = \frac{1}{2} \sum_{i=1}^n \vert \haf{\mathit{PAVA}}(X_i) - \overline{Y} \vert. \]

We therefore have two cases. If 
\begin{equation*}
2\lambda \leq \sum_{i=1}^n \vert \haf{\mathit{PAVA}}(X_i) - \overline{Y} \vert,
\end{equation*}

then we can find solutions to \eqref{Bmin} and \eqref{Amin} for which $A_\lambda \leq \overline{Y} \leq B_\lambda$, producing a minimiser for \eqref{lambdaop}.

Otherwise, \eqref{Bmin} and \eqref{Amin} require that $A > \overline{Y}$ and $B < \overline{Y}$. Hence, there are no solutions to our minimisation with $A<B$.

For a solution on the boundary, we require $A = B$, so
\[L_\lambda(f_{>A, <B}) = \frac{1}{2}\sum_{i=1}^{n} (Y_i - A)^2. \]

This is clearly minimised by $A_\lambda = B_\lambda = \overline{Y}$.

Corollary~\ref{otherzero} follows from the fact that from properties of the PAVA, 
\[\arg\max_{m} \left\vert \sum_{i=1}^m \left( Y_{\pi(i)} - \overline{Y} \right)  \right\vert = \max \left\{ m \textrm{ : } \haf{\mathit{PAVA}}(X_{\pi(m)}) < 0 \right\}.\]

\end{proof}

%%%%%%%%%%%%%%%%%%%%%%%%%%%%%%%%%%%%%%%%%%%%%%%%%%%%%%%%%%%%%%%%%%%%%%%%%%%%%%%%%%%%%%%%%%%%%%%%%%%%%%%%%%%%%%%%%%%%%%%%
\subsection*{Proof of Theorem~\ref{converges}}
\begin{proof}
If we go through the covariates in a pre-determined order, then we can apply a theorem proved in \citet{tseng}. However, the proof is simplified in the case where we go through the variables in a random, independent order with each iteration, which we will show now.

Because we are doing repeated minimisations, for all $m$, $L_\lambda(f^{m+1}) \leq L_\lambda(f^m)$. Moreover, $L_\lambda$ is bounded below by 0. Therefore, $L_\lambda(f^m)$ must converge monotonically in probability as $m$ increases.

Now, choose any $\delta_1 > 0$. We define $A_m$ as the event that there exists $k \in \{\seq{1}{p}\}$ and $g^m: \R \rightarrow \R$ such that 
\[L_\lambda(f_1^m, \ldots, f_p^m) - L_\lambda(f_1^m, \ldots, f_{k-1}^m, g^m, f_{k+1}^m, \ldots, f_p^m) \geq \delta_1,\] the event that we can improve on the current fit by at least $\delta_1$ by changing only one component estimate.

Now, with the $m+1$th iteration, we have an equal probability of picking any one of $p$ covariates as the first fitted of our new backfitting cycle, and hence
\[\textrm{Prob} \left(L_\lambda(f^m) - L_\lambda(f^{m+1}) \geq \delta_1\right) \geq \textrm{Prob} (A_m) / p.\]

But $L_\lambda(f^m)$ converges in probability, so $\textrm{Prob}(A_m) \rightarrow 0$ for all $\delta_1 > 0$.

$L_\lambda$ is continuously differentiable in the interior of $\oplus \mathcal{F}_k$. The set $f \in \oplus \mathcal{F}_k$ such that $L_\lambda(f) \leq L_\lambda(f^{0})$ is closed, and compact. Therefore, the above implies that for any $\eta > 0$, $\delta_2>0$, there exists $M_1$ such that for all $m > M_1$, the subdifferential of $L_\lambda$ at $f^m$ contains a plane that is within $\delta_2$ of zero with probability at least $1-\eta$.

Therefore, by considering sufficiently small values of $\delta_2$, this implies that for all $\eta > 0$, $\delta_3>0$, there exists $M_2$ such that for all $m > M_2$, there exists with probability at least $1-\eta$, another sum of functions $\widetilde{f}^m \in \oplus \mathcal{F}_k$ satistifying $\Vert \widetilde{f}^m - f^m \Vert < \delta_3$, at which $L_\lambda$ contains the zero plane in its subdifferential. 

Since $L_\lambda$ is convex, $\widetilde{f}^m$ is a global minimiser of $L_\lambda$. Taking $\eta, \delta_3$ to zero, we see by continuity of $L_\lambda$ that $L_\lambda(f^m)$ converges in probability to the global minimum.

In addition, this implies that if the global minimum is unique, and so equals $\widetilde{f}^m\quad\forall m$, $f^m$ must converge to it.

\end{proof}

\subsection*{Proof of Theorem~\ref{decomp}}

\begin{proof}
For simplicity, consider only the univariate case. Assume further for simplicity of notation, by permuting the observations if neccessary, that the covariate is sorted in that $X_1 < \ldots < X_n$, for $i = \seq{1}{n-1}$.

Let $\mathcal{G}$ be the set of pairs of monotonically increasing and monotonically decreasing functions $g, h$, with mean zero, with the constraint that in each interval at most one function of the two changes. Hence, with ordered covariate observations, either $g(X_{i+1}) = g(X_{i})$ or $h(X_{i+1}) = h(X_{i})$.

Observe that for any right continuous mean zero step function $f$, any pair of monotonically increasing and monotonically decreasing mean zero functions $g, h$ satisfying $g+h \equiv f$ can only minimise $\Delta{g} + \Delta{h}$ among such functions if $(g, h) \in \mathcal{G}$. Otherwise, if the constraint is broken for $i$, defining 
\begin{align*}
 \widetilde{g}(x) &= g(x) - \min\left(g(X_{i+1}) - g(X_{i}), h(X_{i}) - h(X_{i+1})\right)  I_{\{x > X_{i+1}\}} \\
\widetilde{h}(x) &= h(x) + \min\left(g(X_{i+1}) - g(X_{i}), h(X_{i}) - h(X_{i+1})\right)  I_{\{x > X_{i+1}\}} ,
\end{align*}
gives $\widetilde{g} + \widetilde{h} \equiv g + h$, and $\Delta(\widetilde{g}) + \Delta(\widetilde{h}) = \Delta(g) + \Delta(h) - 2 \min\left(g(X_{i+1}) - g(X_{i}), h(X_{i}) - h(X_{i+1})\right)$.

Therefore solutions to \eqref{decompop} must lie within $\mathcal{G}$.

Now, it is trivial to see that the decomposition in Definition~\ref{decompdef} maps right continuous step functions with mean zero to pairs in $\mathcal{G}$. Two such step functions have the same decomposition if and only if they are equal at all knot points, and so, are for our purposes equivalent. Summation is an inverse with these spaces, since any such step function can be constructed as the sum of its decomposed functions, and we can produce any element in $\mathcal{G}$ by from a right continuous mean zero step function by decomposing its sum. Thus, Definition~\ref{decompdef} gives us a one to one map between the feasible set of \eqref{totalop} and a set that contains all solutions of \eqref{decompop}.

Let $f$ be any right continuous step function with mean zero, and let $(f^+, f^-) \in \mathcal{G}$ be its decomposition. By construction, 
\begin{align*}
 \Delta(f^+) + \Delta(f^-) &= \sum_{i=1}^{n-1} (f(X_{i+1}) - f(X_{i}))I_{\{f(X_{i+1}) - f(X_{i}) > 0\}} + (f(X_{i}) - f(X_{i+1}))I_{\{f(X_{i+1}) - f(X_{i}) < 0\}} \\
&= \sum_{i=1}^{n-1} \vert f(X_{i+1}) - f(X_{i}) \vert = \Delta(f).
\end{align*} 

Hence, 
\begin{align*}
 M_\lambda \left(f^+, f^-\right) &= \frac{1}{2} \norm{Y - f^+(X) - f^-(X)}^2 + \lambda ( \Delta(f^+) + \Delta(f^-)) \\
&= \frac{1}{2} \norm{Y - f(X)}^2 + \lambda  \Delta(f) = L_\lambda \left(f\right).
\end{align*}

Therefore, minimising $L_\lambda$ means the corresponding decomposed functions must minimise $M_\lambda$, and vice versa. Hence, the decomposition/summation transformations give a one to one map between solutions to \eqref{decompop} and \eqref{lambdaop}.

The case of multiple covariates can be dealt with by applying the some argument to each covariate in turn.
\end{proof}
% 
%  \bigskip
%  \begin{center}
%  {\large\bf SUPPLEMENTAL MATERIALS}
%  \end{center}
%  
%  \begin{description}
%  
%  
%  \item[R-package for Liso:] R-package LISO containing code to perform the methods described in the article. (GNU zipped tar file)
% 
%  \end{description}

\bibliographystyle{apalike}
\bibliography{liso.bib} 
\end{document}